\documentclass{moriond}

\usepackage{color}
\usepackage{xspace}
\usepackage{url}
\usepackage{wrapfig,rotating}
\usepackage{sidecap}
\usepackage{amsmath}
\usepackage{amssymb}
\usepackage{mathrsfs}
\usepackage{lineno}
\usepackage{color}
\usepackage{overpic}

\newcommand{\ljets}{$\ell +$jets\xspace}
\newcommand{\dzero}     {D0\xspace}

\newcommand{\ttbar}{\ensuremath{t\bar{t}}\xspace}
\newcommand{\mm}       {\mathrm}
\newcommand{\mTT}{\ensuremath{m_{t\bar{t}}}\xspace}

\newcommand{\ptt}{\ensuremath{p_{T}^{\mathrm{top}}}\xspace}

\newcommand{\aetat}{\ensuremath{|y^{\mathrm{top}}|}\xspace}

\newcommand{\afb}{\ensuremath{A_{{\footnotesize FB}}^{t\bar{t}}}\xspace}
\newcommand{\afbl}{\ensuremath{A_{\mbox{{\footnotesize FB}}}^{\mm{lep}}}\xspace}

\newcommand{\mcatnlo}   {\textsc{mc@nlo}\xspace}

\def\Journal#1#2#3#4{{#1} {\bf #2}, #3 (#4)}

\def\PLB{{\em Phys. Lett.}  B}
\def\PRL{\em Phys. Rev. Lett.}
\def\PRD{{\em Phys. Rev.} D}

\def\EPJ{{\em Eur. Phys. J.} C}

\def\be{\begin{equation}}
\def\ee{\end{equation}}
\def\bea{\begin{eqnarray}}
\def\eea{\end{eqnarray}}

\newcommand{\Photo}{\includegraphics[height=35mm]{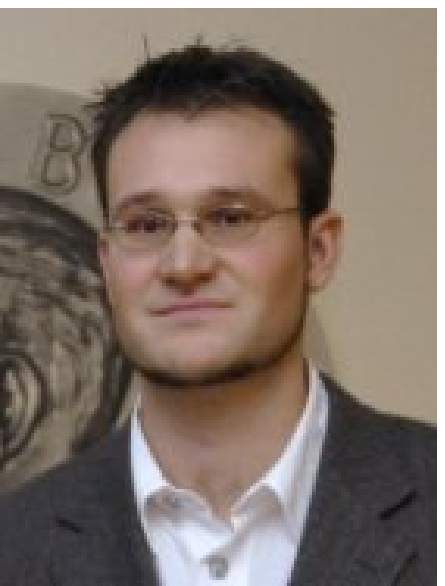}}

\begin{document}
\vspace*{4cm}
\title{TOP QUARK PHYSICS AT THE TEVATRON}


\author{ A.W. Jung }

\address{Fermilab, Particle Physics Division, Kirk Rd \& Pine St,\\
Batavia, 60510, IL, USA\\
Preprint: FERMILAB-CONF-15-208-PPD}

\maketitle\abstracts{
An overview of recent top quark measurements using the full Run II data set of CDF or \dzero at the Tevatron is presented. Results are complementary to the ones at the LHC. Recent measurements of the production cross section of top quarks in strong and electroweak production and of top quark production asymmetries are presented. The latter includes the new measurement of the \ttbar production asymmetry by \dzero in the dilepton decay channel. Within their uncertainties the results from all these measurements agree with their respective Standard Model expectation. Finally latest updates on measurements of the top quark mass are discussed, which at the time of the conference are the most precise determinations.
}

\section{Introduction}
\label{toc:intro}
The top quark is the heaviest known elementary particle and was discovered at the Tevatron $p\bar{p}$ collider in 1995 by the CDF and \dzero collaboration \cite{top_disc1,top_disc2} with a mass around $173~\mathrm{GeV}$. The Tevatron continues to produce complementary and similarly precise results than those at the LHC. At the Tevatron the production is dominated by the $q\bar{q}$ annihilation process, while at the LHC the gluon-gluon fusion process dominates. The top quark has a very short lifetime, which prevents the hadronization process of the top quark. Instead bare quark properties can be observed.\\
The measurements presented here are performed using either the dilepton ($\ell \ell$) final state or the lepton+jets (\ljets) final state. Within the \ljets~final state one of the $W$ bosons (stemming from the decay of the top quarks) decays leptonically, the other $W$ boson decays hadronically. For the dilepton final state both $W$ bosons decay leptonically. The branching fraction for top quarks decaying into $Wb$ is almost 100\%. Jets originating from a $b$-quarks are identified ($b$-tagged) by means of multi-variate methods employing variables describing the properties of secondary vertices and of tracks with large impact parameters relative to the primary vertex. Details on a typical \ttbar event selection, applied requirements to reduce background contributions, and the determination of the sample composition can be found in Ref.~\cite{d0note_diff}.

\section{Single top quark and top quark pair production}
CDF and \dzero conclude their measurement program of single top quark production and final results are discussed here. \dzero performs a simultaneous measurement of the $s$- and $t$-channel electroweak single top-quark production cross sections \cite{d0_stop}. Three multivariate analyses are used to separate the signal from the background. A two-dimensional discriminant based on the combination of the three methods is used to measure the $s$-, $t$- and $s+t$-channel cross sections in one analysis.
\begin{figure}[ht]
   \centering
\begin{overpic}[width=0.55\columnwidth]{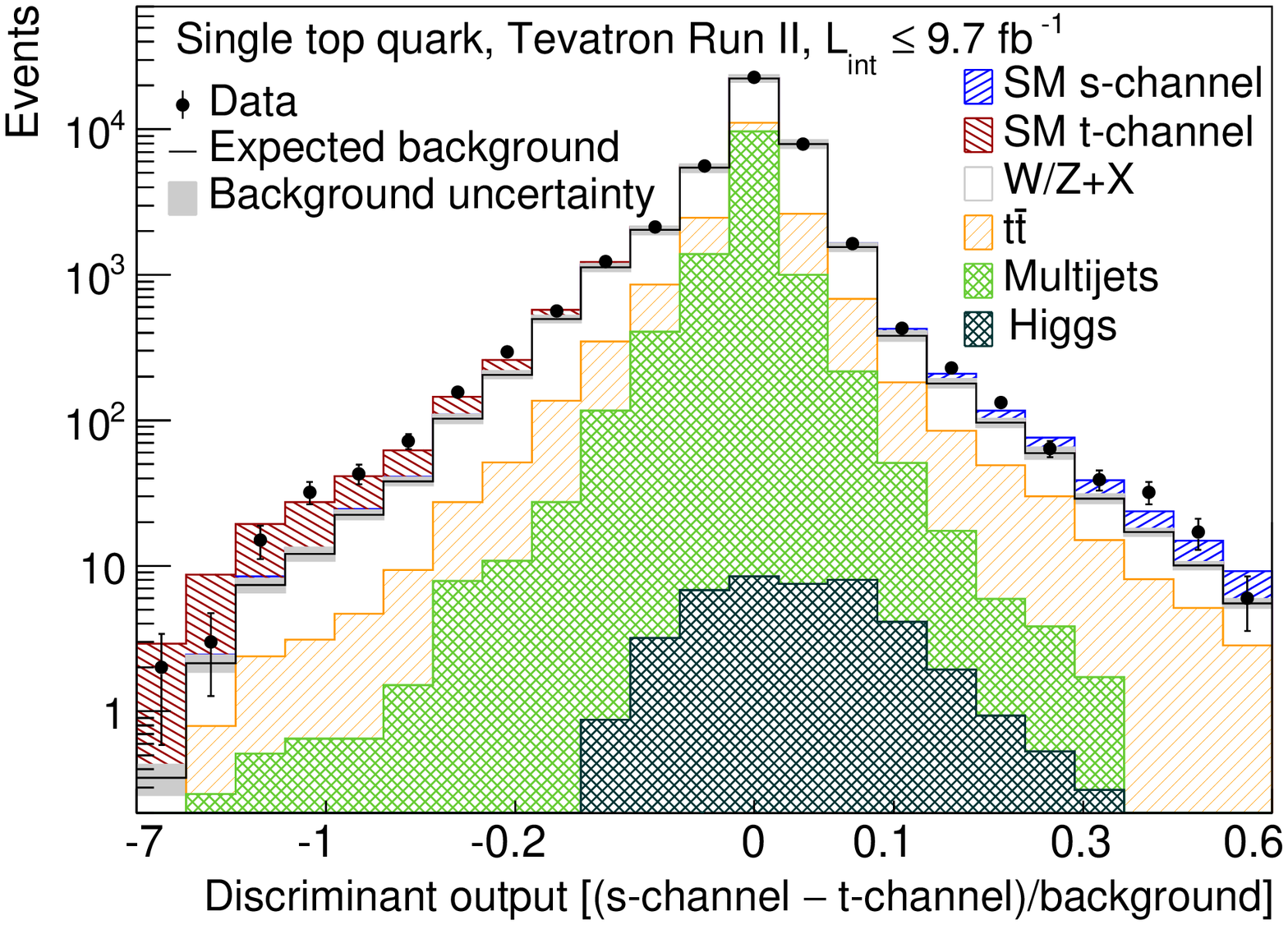}
\put(-7,67){\large\textsf{\textbf{(a)}}} \end{overpic} \hspace{5pt}
\begin{overpic}[width=0.35\textwidth]{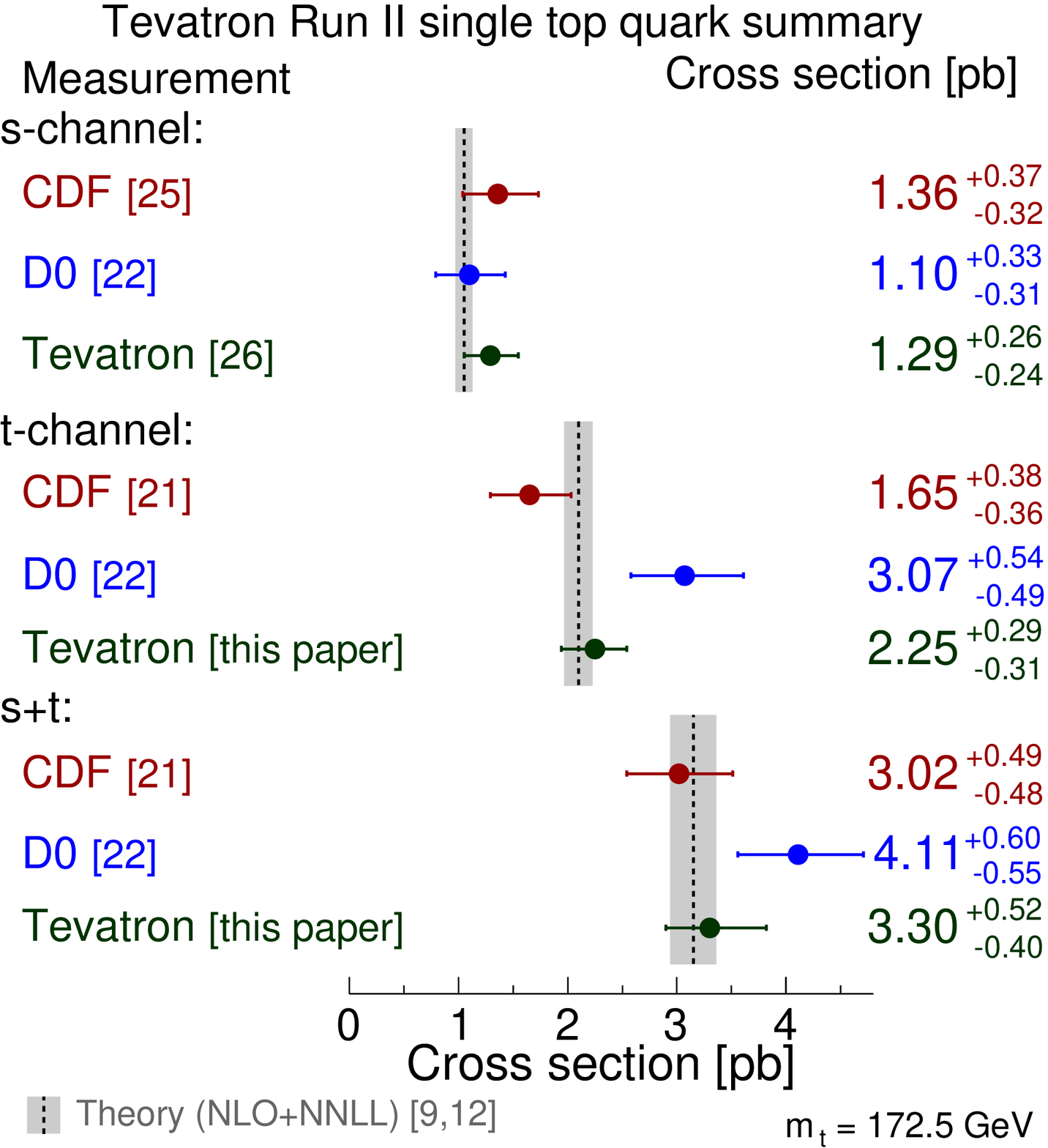}
\put(-9,97){\large\textsf{\textbf{(b)}}} \end{overpic}
 \protect\caption{\label{fig:tev_stop_obs} Measurements of (a) single top quark production cross sections. Distribution of the (b) discriminant output comparing data with signal single top $s$- and $t$-channel contributions as well as various background contributions, all normalized to their expected value. Background uncertainties are indicated by the shaded band on top of the sum of the expected background contributions.}
 \end{figure}
Integrating over the $s$-channel or $t$-channel distribution, the $t$-channel or $s$-channel cross section is measured, respectively. No assumptions are made on the relative contribution of the $s$- or $t$-channel. CDF follows a similar strategy to separate signal and background by employing multivariate analyses in order to measure the individual single top production cross sections in various decay channels \cite{cdf_stop_ljets,cdf_stop_met,cdf_stop_combi,cdf_stop_spt}. Figure \ref{fig:tev_stop_obs}(a) shows the combined \dzero and CDF discriminant output used to extract the $t$- and $s+t$-channel cross sections with the signal cross sections and the various background contributions. Earlier \dzero and CDF performed a combination of results on $s$-channel production cross sections, which yields a cross section of \mbox{$\sigma_{s\mm{-ch.}} = 1.29^{+0.26}_{-0.24}$ pb} with a significance of 6.3 s.d.~corresponding to the first observation of $s$-channel single top quark production \cite{tev_stop}. A summary of single top $s$-, $t$- and $s+t$-channel cross section results at \dzero and CDF is given in Figure \ref{fig:tev_stop_obs}(b). All measurements are in good agreement with the latest theoretical calculations. A direct limit on the CKM matrix element $V_{tb} > 0.92$ at 95\% confidence level is derived from the combined $s+t$-channel cross section measurement.

\subsection{Top quark pair production}
\begin{figure}[ht]
\centering
  \includegraphics[width=0.85\columnwidth,angle=0]{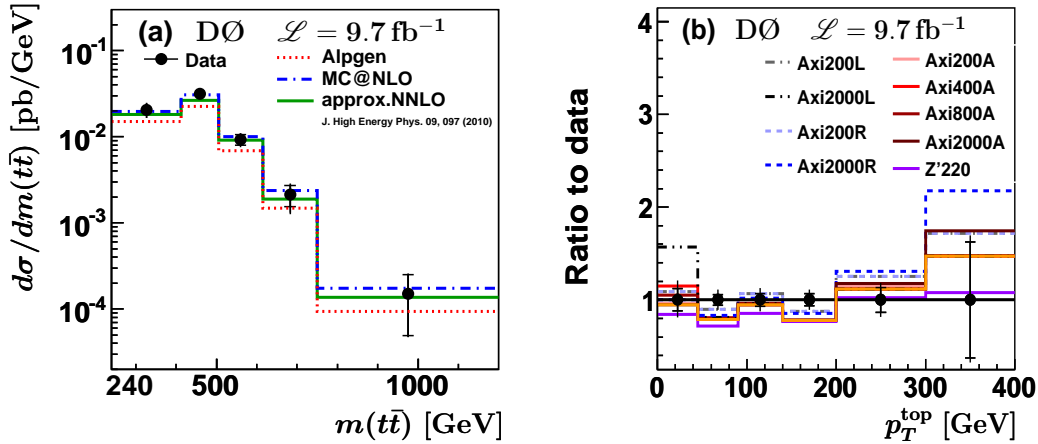}
\protect\caption{\label{fig:topxsec_axiModels1} Differential cross section data as a function of (a) \mTT compared with expectations from QCD. The inner error bar represents the statistical uncertainty, whereas the outer one is the total uncertainty including systematic uncertainties. Ratio of (b) differential cross section distributions as a function of \ptt to various benchmark models of axigluon contributions to the \ttbar production cross section are shown.}
\end{figure}

\dzero uses events in the lepton+jets decay channel \cite{d0note_diff} to study differential top quark cross sections as a function of $p_T$ (\ptt), the absolute value of the rapidity $|y|$ (\aetat), as well as the invariant mass of the \ttbar pair, \mTT. The most direct constraint for contributions of new physics is set by the \mTT distribution, which is sensitive to the production of resonances decaying into top quarks, like a $Z'$. To identify the top quarks, a kinematic reconstruction, which takes into account experimental resolutions, is performed. All possible permutations of objects are considered, while preferentially assigning $b$-tagged jets to $b$-quarks and the chosen solution is the one with the smallest $\chi^2$. The differential cross section is shown in Figure \ref{fig:topxsec_axiModels1}(a) as a function of the invariant mass \mTT compared to various predictions. Figure \ref{fig:topxsec_axiModels1}(b) shows the ratio of the differential cross section as a function of \ptt to the approximate NNLO and various axigluon models \cite{axigluon} that could alter the production of \ttbar events. Data agrees with the SM predictions. Models implementing heavy axigluon masses are already in tension with existing data from the Tevatron and the LHC, but it is especially the low mass region where the Tevatron data adds sensitivity. The low-mass $Z'$ model shows significant tension to the data in all three differential distributions.\\

\section{Top quark production asymmetries}
\label{toc:angular}
The different initial state makes measurements of angular correlations in $t\bar{t}$ events, such as production asymmetries, complementary between the Tevatron and the LHC. 
\begin{figure}[ht]
\begin{overpic}[width=0.55\columnwidth]{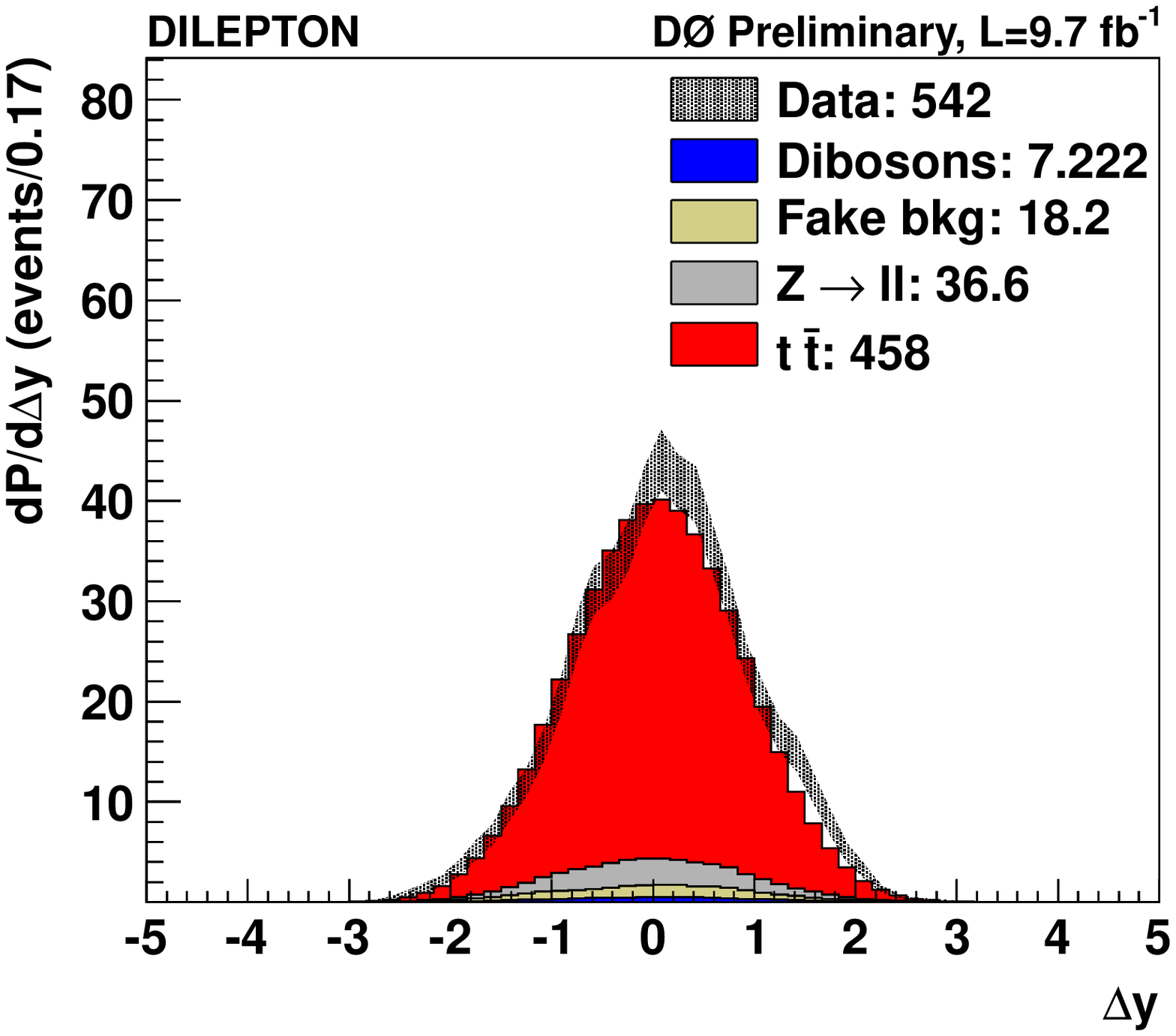}
\put(3,75){\large\textsf{\textbf{(a)}}} \end{overpic}
\begin{overpic}[width=0.35\textwidth]{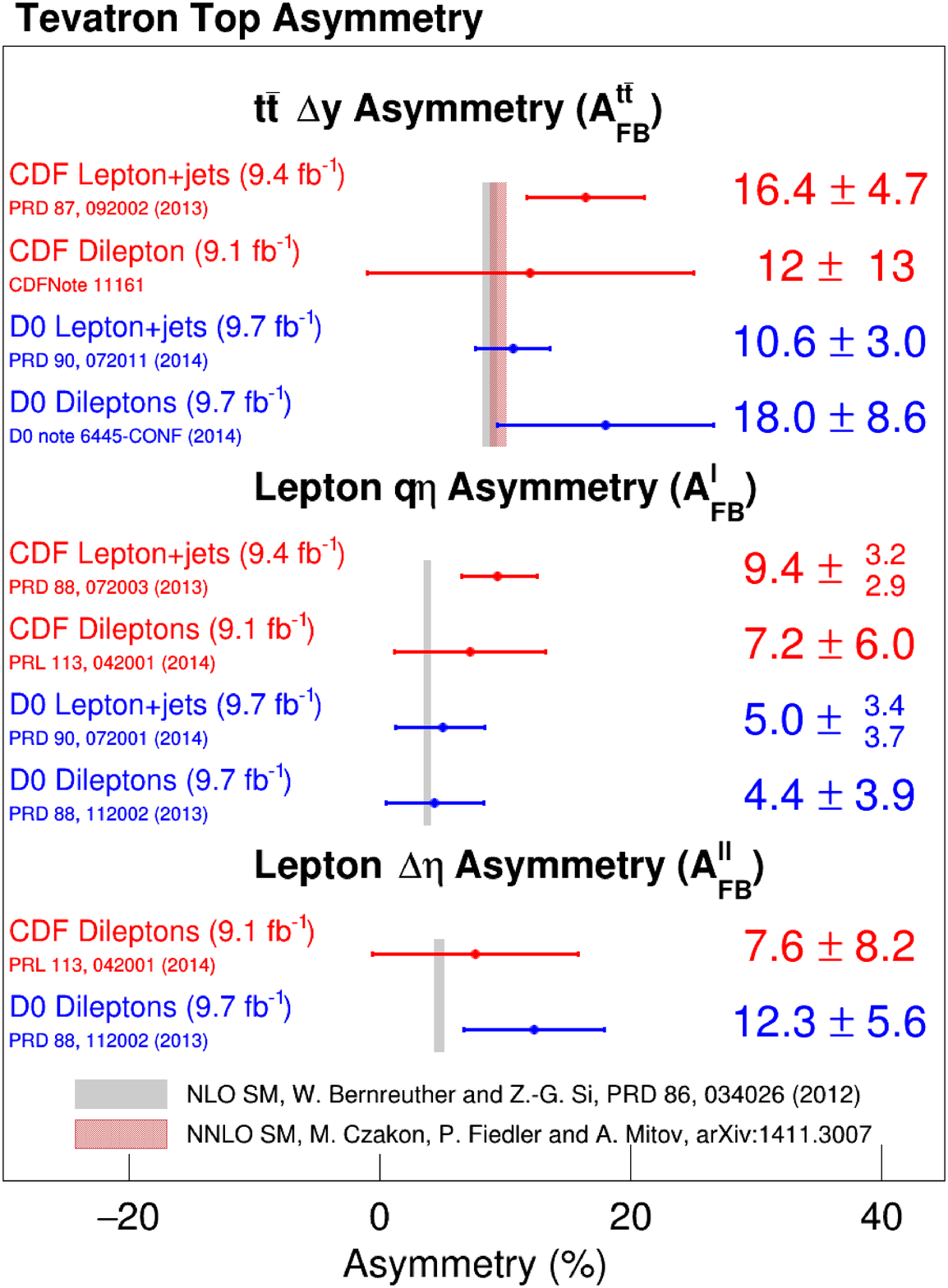}
\put(-9,92){\large\textsf{\textbf{(b)}}} \end{overpic}
 \protect\caption{\label{fig:d0ll} 
 The (a) $\Delta y$ distribution for selected events compared to the expectation from \mcatnlo and various background contributions. The use of the matrix elements technique reflects in correlated reconstructed $\Delta y$ values, hence data is indicated by the shaded black band. Summary (b) of \afb measurements at the Tevatron compared to SM predictions, more details see text.}
 \end{figure}
Experimentally, there are two approaches to measure these asymmetries: Either only a final-state particle, e.g. a lepton (`lepton-based asymmetries') is reconstructed or top quarks are fully reconstructed using a kinematic reconstruction. The forward-backward asymmetry \afb at the Tevatron measures $\Delta y = y_t - y_{\bar{t}}$, and employing this quantity the production asymmetry is defined as:
\begin{equation}
A_{\mbox{{\footnotesize FB}}}^{t\bar{t}} = \dfrac{N(\Delta y >0) - N(\Delta y <0)}{N(\Delta y >0) + N(\Delta y <0)}
\end{equation} \\
Calculations at NLO QCD including electroweak corrections \cite{bernSi} predict \afb$=0.088 \pm 0.005$ and \afbl$=0.038 \pm 0.005$. Recently predictions at NNLO+NNLL pQCD including electroweak corrections became available with a predicted value of \afb$= 0.095 \pm 0.007$ \cite{mitov}. Latest updates provide a prediction at approximate N$^3$LO pQCD including electroweak corrections with a predicted value of \mbox{\afb$= 0.100 \pm 0.006$ \cite{kidonakis}}.\\
The most recent experimental update is by \dzero, which presented a measurement of the fully reconstructed top quark asymmetry in the dilepton decay channel \cite{d0_ttbar_ll_afb}. The measurement employs the full Run II data set corresponding to an integrated luminosity of 9.7/fb. Events with at least two jets, two high momentum and isolated electrons or muons or one high momentum isolated electron and muon are selected together with requiring a large missing transverse energy corresponding to the non-detected neutrinos of the leptonic $W$ boson decay. To fully reconstruct the \ttbar event a matrix element technique is applied, which calculates a likelihood of all the possible combinations when assigning reconstructed quantities to parton level \ttbar quantities.\\
Figure \ref{fig:d0ll}(a) shows the $\Delta y = y_t - y_{\bar{t}}$ distribution for the selected data events compared to the signal expectation from \mcatnlo and various background contributions. The measurement is corrected for detector effects to the parton level. If the measurement is interpreted as a test of the SM the measurement yields \afb$=0.180 \pm 0.061\,\mm{(stat.)} \pm 0.032\,\mm{(syst.)}$. Due to the unknown top quark polarization an additional model uncertainty of 5.1\% applies once the measurement is interpreted as a search for contributions of new physics. \\
A summary of \afb and \afbl measurements at the Tevatron is given in Figure \ref{fig:d0ll}(b). For measurements of \afb the deviations from the SM predictions got smaller with the \dzero measurement \cite{d0_ttbar_afb} early last year employing the full data set. The recent NNLO+NNLL pQCD calculations are in agreement with the \dzero data. CDF results with the full data set are showing deviations at the 1 to 2 s.d.~level, especially the differential \afb measurement \cite{cdf_ttbar_afb} shows larger deviations. It should be noted that efforts toward a Tevatron combination of \afb and \afbl measurements are currently ongoing.

\section{Top Quark Mass}
\label{toc:mass}
A large variety of other measurements of top quark properties at the Tevatron exists to date \cite{d0_web,cdf_web} and is not discussed in detail here. Measurements of the top quark mass use different experimental techniques in order to extract the top quark mass. The latest update is a measurement by \dzero applying the leading order Matrix Element method based on an event-by-event probability. All top quark mass measurements applying standard methods are dominated by systematic uncertainties. Furthermore there is an additional theoretical uncertainty of about $1$ GeV, which originates from the implementation of the quark mass in the MC employed to measure the top quark mass, aka pole vs $\bar{MS}$ mass discussion. Strategies to overcome the limitations in terms of experimental and theoretical uncertainties are already pursued and will become more important for the upcoming run of the LHC.\\
\begin{figure}[ht]
\begin{overpic}[width=0.45\columnwidth]{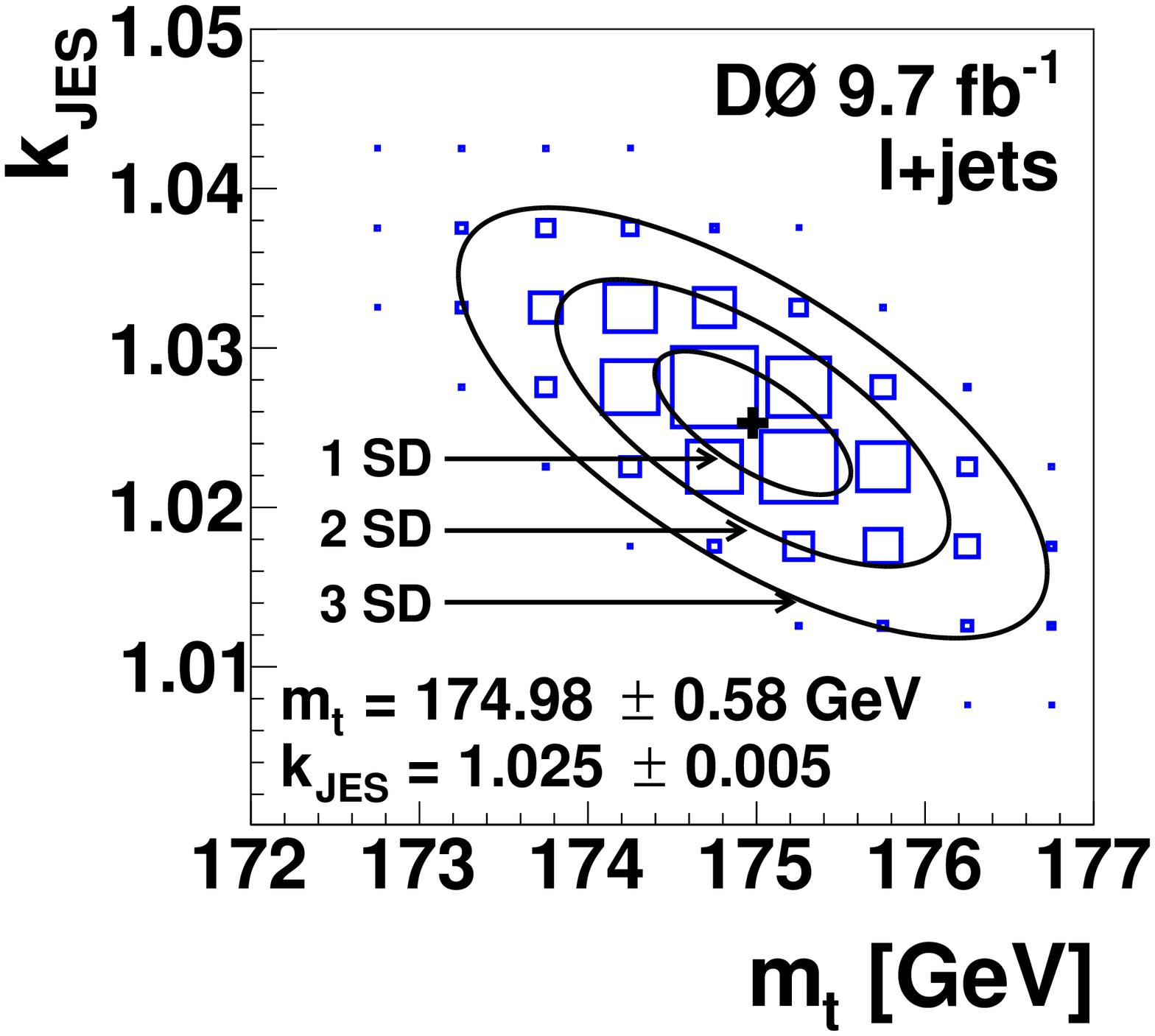}
\put(-5,85){\large\textsf{\textbf{(a)}}} \end{overpic} \hspace{7pt}
\begin{overpic}[width=0.45\textwidth]{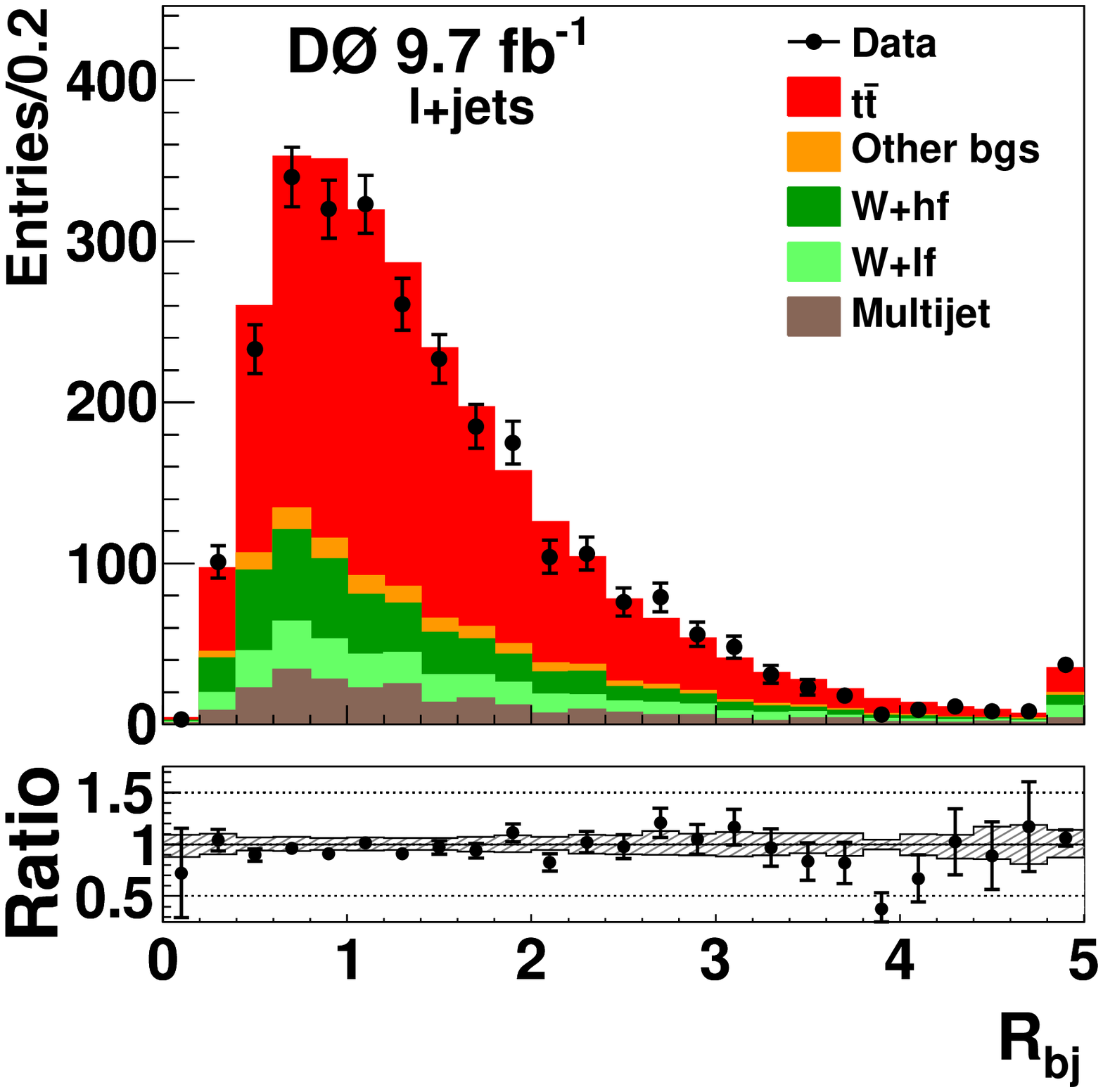}
\put(-8,85){\large\textsf{\textbf{(b)}}} \end{overpic}
  \protect\caption{\label{fig:mass_cms_d0} The (a) two-dimensional likelihood as a function of the top quark mass and the in-situ calibration factor as measured by \dzero. The (b) ratio of the $p_T$ of $b$-tagged jets to light quark jets originating from the hadronic $W$ decay. Systematic uncertainties are indicated by the hashed band in the Data to MC ratio at the bottom of the histogram.}
 \end{figure}
Currently the single most precise measurement of any experiment is done by \dzero in the \ljets decay channel \cite{d0_mass} using the full Run II data set. It employs the so-called matrix element method (ME), which calculates an event-by-event probability to match the \ttbar final state in the \ljets decay channel to the observed reconstructed objects taking into account detector resolutions. 
\begin{wrapfigure}{r}{0.505\textwidth}
\centerline{\includegraphics[width=0.45\textwidth]{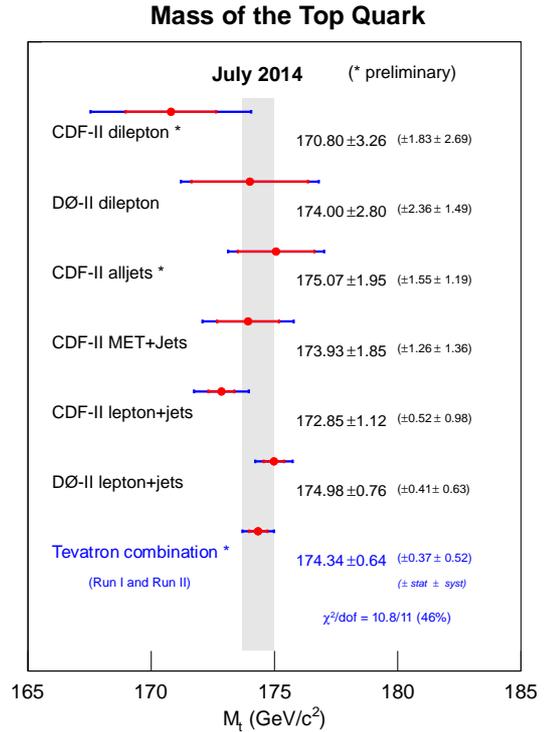}}
\caption{\label{fig:charge} \label{fig:world} Most precise input measurements to the latest Tevatron top quark mass combination. Measurements are in good agreement with each other as indicated by the $\chi^2$ value.}
\end{wrapfigure}%
The hadronic decay of one of the $W$ bosons allows to constrain the jet energy scale in-situ from data. Figure \ref{fig:mass_cms_d0}(a) shows the two-dimensional likelihood as a function of the top quark mass and the in-situ calibration factor. It yields a mass of $m_{t} = 174.98 \pm 0.41\, (\mm{stat.}) \pm 0.64\,(\mm{sys.\,+\,JES})~\mm{GeV}$, corresponding to a total relative uncertainty of 0.43\%. Figure \ref{fig:mass_cms_d0}(b) shows the ratio $R_{\mm{bj}}$ of the $p_T$ of $b$-tagged jets to light quark jets originating from the hadronic $W$ decay, which is sensitive to the $b$ jet energy scale. Using MC template distributions for various true values of the $b$ jet energy scale a cross check is performed, which yields $R_{\mm{bj}} = 1.008 \pm 0.0195\,(\mm{stat.}) ^{+0.037}_{-0.031}\,(\mm{syst.})$ in good agreement with the expectation of unity \cite{PRDmass_ljets}.\\ 
The latest Tevatron combination of top quark mass measurements \cite{mt_tev} by CDF and \dzero combines all existing measurements. The combined top quark mass is $m_{t} = 174.34 \pm 0.64\,(\mm{stat.\,+\,sys.\,+\,JES})~\mm{GeV}$, corresponding to a total relative uncertainty of 0.37\%.\\
Direct measurements of the top quark mass are becoming ever more precise and provide a stringent self-consistency test of the SM by correlating $m_t$ versus $m_W$. Together with the measurement of the mass of the recently discovered Higgs boson \cite{higgs1,higgs2} this is a strong self-consistency test of the SM \cite{gfitter}. Furthermore the stability of the electroweak vacuum can be studied \cite{vacuum}.

\section{Conclusions}

Various recent measurements of top quark properties at the LHC and at the Tevatron are discussed. CDF and \dzero conclude their measurement program of single top quark production and final results were presented at this conference. Pair production cross sections provide stringent tests of the SM calculations and do not show any hint for deviations from the SM. Direct measurements of the top quark mass are becoming ever more precise and provide a stringent self-consistency test of the SM and new insights into the question of the stability of the electroweak vacuum. Unlike in the past the measurements of \afb and \afbl basically agree with the latest SM predictions at the 1 to 2 s.d.~level. Studies on combinations of \afb and \afbl at the Tevatron are currently ongoing. All of the presented results in terms are in good agreement with the Standard Model expectations and do not show any hints for new physics.

\section*{Acknowledgments}

The author thanks the organizers of the $50^{\mm{th}}$ Moriond conference for the invitation and for the hospitality of the conference venue.

\end{document}